\documentclass[twocolumn]{aastex63}
\usepackage{amsmath}
\UseRawInputEncoding
\submitjournal{AJ}

\shorttitle{Modulation in V838 Cyg}
\shortauthors{Li et al.}

\begin{document}

\title{Two Different Weak Modulations in ab-type RR Lyrae Variable V838 Cyg, and Potential Influence of Metal Abundance on Blazhko Modulation}

\correspondingauthor{L.-J. Li}
\email{lipk@ynao.ac.cn}

\author{Li, L.-J.}
\affiliation{Yunnan Observatories, Chinese Academy of Sciences,
P.O. Box 110, Kunming, 650216, PR China}
\affiliation{Key laboratory of the structure and evolution
of celestial objects, Chinese Academy of Sciences, P.O. Box 110, Kunming,
650216, PR China}

\author{Qian, S.-B.}
\affiliation{Yunnan Observatories, Chinese Academy of Sciences,
P.O. Box 110, Kunming, 650216, PR China}
\affiliation{Key laboratory of the structure and evolution
of celestial objects, Chinese Academy of Sciences, P.O. Box 110, Kunming,
650216, PR China}
\affiliation{University of Chinese Academy of Sciences, No.1 Yanqihu East Rd, Huairou District, Beijing, 101408, PR China}

\author{Shi, X.-D.}
\affiliation{Yunnan Observatories, Chinese Academy of Sciences,
P.O. Box 110, Kunming, 650216, PR China}
\affiliation{Key laboratory of the structure and evolution
of celestial objects, Chinese Academy of Sciences, P.O. Box 110, Kunming,
650216, PR China}

\author{Zhu, L.-Y.}
\affiliation{Yunnan Observatories, Chinese Academy of Sciences,
P.O. Box 110, Kunming, 650216, PR China}
\affiliation{Key laboratory of the structure and evolution
of celestial objects, Chinese Academy of Sciences, P.O. Box 110, Kunming,
650216, PR China}
\affiliation{University of Chinese Academy of Sciences, No.1 Yanqihu East Rd, Huairou District, Beijing, 101408, PR China}



\begin{abstract}

Noting the weakest modulation and relatively high metal abundance of the ab-type RR Lyrae star V838 Cyg, we collected the photometric data of this star from several sky surveys to carry out an in-depth analysis. The $O-C$ diagram shows that the pulsation period of V838 Cyg increases linearly over a long timescale. In a reanalysis of the high-precision $Kepler$ data, we confirmed the modulation with a period of $59.45\pm0.07$ days found by Benk\H{o} et al., (2014), and also found an additional weak modulation with a longer period ($840\pm21$ days). After a series of analyses, we incline to the view that the mechanisms causing the two modulations are different: the former is more similar to the typical Blazhko effect, while the mechanism leading to the latter may be an extrinsic factor. We also collected and compared the modulation and physical parameters of other Blazhko RR Lyrae stars from several works in the literature, and find that there is a potential negative correlation between the modulation amplitude (or upper limit of amplitude) and the metal abundance. We infer that the relatively high metal abundance will promote convection in the outer stellar atmosphere, and then inhibit those factors (turbulence, shock wave, etc.) that may cause Blazhko modulation. Future observations and research work can be carried out with reference to this viewpoint. We also introduce the moir\'{e} effects that appear in the $Kepler$ long-cadence light curves and their possible interference in the previous analyses.

\end{abstract}




\section{Introduction} \label{sec:intro}

RR Lyrae stars are short-period pulsating variables in the evolutionary stage of the horizontal branch (HB), generally with pulsation periods of 0.2 - 1.2 days and pulsation amplitudes of 0.2 - 1 mag in the $V$ band. Because of these pulsating characteristics, they are easy to identify in observations. In fact, RR Lyrae stars have been discovered for more than 100 years, and have a long history of research and systematic achievements \citep{2004rrls.book.....S}. In the recently released Gaia DR3 catalog, scientists have provided up to 270,000 RR Lyrae stars located in the Milky Way and nearby galaxies \citep{2022yCat.1358....0G}. Due to their wide distribution and large number, RR Lyrae stars are often used as probes to study the chemical, evolutionary, and dynamical properties of old low-mass stars in the Milky Way \citep{2004rrls.book.....S}.

However, in the field of RR Lyrae stars, there has always been an important feature of observations and research, that is, the Blazhko effect \citep{1907AN....175..325B} and the dominant physical mechanism behind it. Most RR Lyrae stars are single-mode pulsating variables: the ab type of RR Lyrae stars (RRab stars) dominated by the fundamental mode or the c type (RRc stars) dominated by the first overtone mode. In the observations, the light curves of some RR Lyrae stars show good stability, but a considerable number of stars show phenomena similar to the amplitude and frequency modulations in radio technology \citep{2011MNRAS.417..974B}, which are specifically manifested as periodic or quasi-periodic changes of pulsation parameters over time with periods from a few days to hundreds of days \citep{2008JPhCS.118a2060K}. With the acquisition of ground-based and space-based photometric data in recent years (especially the $Kepler$ mission, \citealt{2010Sci...327..977B,2010ApJ...713L..79K}), some discoveries have been made from the perspective of observation, including the conclusion that about 40\% - 50\% of the Galactic RRab stars exhibit the Blazhko effect \citep{2009MNRAS.400.1006J,2013ApJ...773..181N,2017MNRAS.466.2602P}. Of course, there are also studies that suggest that more than 90\% of RR Lyrae stars have additional components in the frequency spectrum of the light curves \citep{2018A&A...614L...4K,2021ASPC..529...51K}. This difference in the rate of incidence should be caused by different definitions of the Blazhko effect. As a macroscopic phenomenon, stellar pulsation is affected by a variety of factors throughout the process. When the observational accuracy is high enough, all RR Lyrae stars should be found to have additional oscillation and modulation components. Our viewpoint is to limit the Blazhko effect to those modulations that exhibit periodicity or quasi-periodicity. Otherwise, arbitrarily expanding the definition will cause the research to lose its pertinence.

Several theoretical models have been proposed to explain the Blazhko effect, but no consensus has been reached so far \citep{2015ApJ...802...52B}. Perhaps it is valuable to find more clues and summarize more patterns from observations \citep{2012AJ....144...39L,2020MNRAS.494.1237S}. Some works have been done on the statistical analysis of pulsation parameters and Blazhko modulation parameters. \citet{2005IBVS.5666....1J} studied the relationship between the modulation amplitude and pulsation frequency of Blazhko stars, and found that both the modulation amplitude and the pulsation amplitude increased with decreasing of the pulsation period. 

Metal abundance is one of the most important parameters in stellar evolution. In the study of HB stars in globular clusters, metal abundance is the first factor affecting HB morphology (the 'first parameter', \citealt{1960ApJ...131..598S}). Metal abundance is also very important in the field of RR Lyrae stars. It is associated with absolute magnitude: metal-poor RR Lyrae stars are brighter and more massive \citep{2007A&A...476..779B,2010ApJ...722...79S}. Metal abundance is also related to pulsation characteristics (i.e., the $P_{\rm pul}$ - $\phi_{31}$ - [Fe/H] relation, see \citealt{2021ApJ...912..144M,2022ApJ...931..131M}, and references therein). Some works have also accumulated on whether there is a relationship between metal abundance and the Blazhko effect. \citet{1976ASSL...60..133S} found that the proportion of Blazhko RRab stars increased with $\Delta S$, and pointed out that the Blazhko stars generally have rather low metal abundance. \citet{2004rrls.book.....S} compared the distribution of metal abundance in Blazhko RRab stars with the distribution of bright RRab stars, and concluded that "No trend in prevalence of the Blazhko effect as a function of metallicity is evident in this comparison." \citet{2005AcA....55...59S} studied the Blazhko RRab stars in the Galactic bulge and the Large Magellanic Cloud, and found that the photometric metal abundance distributions of non-Blazhko and Blazhko variables are similar, and the difference in metal abundance cannot explain the rate of incidence of modulation that occurs in the two systems. Based on the studies on RR Lyrae stars in the $Kepler$ field, \citet{2015ApJ...809L..19B} have noticed that there seems to be some relationship between the metal abundance and the amplitude of modulation, but did not carry out further research. \citet{2020MNRAS.494.1237S} found that the possible range of modulation amplitudes decreases with increasing pulsation period and deduced that the Blazhko effect was suppressed in more metal-poor RR Lyrae stars. It can be seen that, from the perspective of observation, researchers do not have a unified view and understanding of the correlation between metal abundance and Blazhko modulation. 

V838 Cyg had been examined at the Maria Mitchell Observatory for about 50 years \citep{1978JAVSO...7...79S}. A total of 70 times of light maximum were available in the GEOS database\footnote{\url{http://rr-lyr.irap.omp.eu/dbrr/}}. Based on these these data and $Kepler$ data, \citet{2011MNRAS.417.1022N} obtained the long-term rate of change of period of V838 Cyg: 0.05(4) day Myr$^{-1}$. At first, V838 Cyg was considered as a non-Blazhko RRab star \citep{2010MNRAS.409.1585B,2011MNRAS.417.1022N}, but \citet{2013ApJ...773..181N} found that this variable star exhibited amplitude and frequency modulations with the lowest range, and identified it as a Blazhko star. In addition, they further pointed out that there is a variation in the modulation period, from 47, 54 to 64 days. \citet{2014ApJS..213...31B} used an independent method to detect this weak amplitude modulation with a period of 59.5 days, and also pointed out that V838 Cyg shows the period doubling phenomenon \citep{2010ApJ...713L.198K,2010MNRAS.409.1244S} and the second radial overtone mode $f_{2}$. In addition, the metal abundance of V838 Cyg is the largest among all Blazhko RRab stars in the $Kepler$ field ([M/H] = -1.01, \citealt{2013ApJ...773..181N}). Noting these characteristics, we use the photometric data collected from different sky surveys to conduct targeted analysis on V838 Cyg, and try to find some relevant indications. In Section \ref{sec:surveys}, we present the photometric data from different sky surveys, including ground-based and space-based data. We describe the detailed analysis in Section \ref{Sec:Analysis}, and provide the corresponding discussions in Section \ref{Sec:4}. We discuss the moir\'{e} patterns in the long-cadence light curves and their influence on and interference in the analyses in Section \ref{Sec:5}. The potential relationship between metal abundance and modulations is discussed in Section \ref{Sec:6}. In Section \ref{Sec:Summary}, we present our summary.

\section{Light curves from sky surveys} \label{sec:surveys}

We collect the photometric data of V838 Cyg from several sky survey projects, including the ground data provided by SWASP and ASAS-SN, and the space data from $Kepler$ and TESS. The former are mainly used to study the long-term change in pulsation period, while the $Kepler$ data with high accuracy are used to study the weak modulation of V838 Cyg \citep{2013ApJ...773..181N,2014ApJS..213...31B}. In addition, the GEOS RR Lyr database provides 70 times of light maximum for V838 Cyg \citep{2007A&A...476..307L}, and we use those published ones for $O-C$ analysis. We introduce these data from different sources in the following subsections.


\subsection{SWASP} \label{sec:surveys:SWASP}

SWASP\footnote{\url{https://www.superwasp.org/}} (SuperWASP, Super Wide Angle Search for Planets) is one of the early sky survey projects to carry out ground-based observations searching for exoplanets. It consists of two robotic observatories, respectively located on the island of La Palma and in South Africa, allowing the observations to cover both the northern and southern hemispheres of the sky. Their observation equipment comprises eight wide-angle cameras, and the effective aperture of each lens is about 20 cm (see \citealt{2010A&A...520L..10B} for more details). The project was carried out from 2004 to 2008, used the transit method to find exoplanets, and was one of the important exploration projects in related fields before the $Kepler$ space mission \citep{2014CoSka..43..500S}. In addition, studies of variable stars and other fields also benefit from SWASP (i.e. \citealt{2011A&A...528A..90N,2011A&A...535A...3S,2014A&A...562A..90S,2015AN....336..981B,2016MNRAS.459.4360L,2017MNRAS.465.3889R}). For V838 Cyg, SWASP provides more than 7600 photometric data points from different cameras. Based on the accuracy of the data, we use the data observed by cameras 141 and 144 for analysis.


\subsection{ASAS-SN} \label{sec:surveys:ASASSN}

The science goal of ASAS-SN\footnote{\url{https://www.astronomy.ohio-state.edu/asassn}} (All-Sky Automated Survey for Supernovae), as its name suggests, is to search for bright supernovae and transients \citep{2014ApJ...788...48S}. This project now has 24 telescopes distributed around the globe, surveying the entire visible sky down to about 18th magnitude. This project provides 384 photometry data in the $V$ band for V838 Cyg \citep{2018MNRAS.477.3145J}. We use these data to study the long-term change in the pulsation period (see Subsection \ref{Sec:Analysis:sub2}).

\subsection{$Kepler$ and TESS} \label{sec:surveys:KeplerTESS}

The goal of the $Kepler$ mission is to discover exoplanets by the transit method \citep{2010Sci...327..977B}, and it opened up a new era in this field. The number of exoplanets discovered by the $Kepler$ project has reached thousands\footnote{\url{http://exoplanet.eu/}}. This mission also greatly promoted research in the field of RR Lyrae stars \citep{2021FrASS...7...81P}. Based on $Kepler$ data, V838 Cyg has been studied by \citet{2010MNRAS.409.1585B,2014ApJS..213...31B} and \citet{2011MNRAS.417.1022N,2013ApJ...773..181N}. In the present paper, we use the rectified data (Quarters 1-16) provided by \citet{2014ApJS..213...31B} and the Quarter 17 long-cadence data taken from MAST (Barbara A. Mikulski Archive for Space Telescopes\footnote{\url{https://mast.stsci.edu/portal/Mashup/Clients/Mast/Portal.html}}) for analysis. 61,183 long-cadence data points in total were used in our analysis.

Similar to $Kepler$, the goal of TESS (Transiting Exoplanet Survey Satellite) is to search for exoplanets near the solar system by using the transit method \citep{2015JATIS...1a4003R}. However, due to the huge field of view, its resolution is very low, only 21 arcseconds/pixel\footnote{\url{https://tess.mit.edu/science/}}. Faint targets, such as V838 Cyg, are easily affected by nearby starlight. We download the $5\times5$ pixel image cutouts of five sectors (Sectors 14, 15, 40, 41, and 53, \citealt{2019ascl.soft05007B}\footnote{\url{https://mast.stsci.edu/tesscut/}}), and obtain the light curves of V838 Cyg in flux by using the following equation:
\begin{equation}
\Delta flux(t) = [\frac{\sum\limits^{9}_{\rm inner=1} flux\_inner(t)}{9}-\frac{\sum\limits^{16}_{\rm outer=1}flux\_outer(t)}{16}], \label{Equ:1}
\end{equation}
where $flux\_inner$ represents the flux in the central $3\times3$ pixels of each cutout, and $flux\_outer$ represents the flux of the 16 pixels around the central nine pixels. Here we need to explain why Equation (\ref{Equ:1}) is used to obtain the light curves. There is a brighter variable star near V838 Cyg (KIC 10789302, identified as a rotating variable, \citealt{2014ApJS..211...24M}), and the difference in position between the two stars is only about 70 arcseconds (only two pixels apart in the frame image). This makes it difficult to produce the light curves of V838 Cyg from the images. We have tried various equations, and Equation (\ref{Equ:1}) probably produces the best results. We ultimately obtained 13,495 data points for analysis, which are available in machine-readable format in a .tar.gz package. It should be mentioned that the flux amplitude of the light curve is not the true amplitude. We use these data to determine the times of light maximum and minimum for $O-C$ analysis.

\section{Analyses} \label{Sec:Analysis}

The analysis methods we adopt are mainly the $O-C$ method, Fourier spectrum method, and corresponding research methods for modulation \citep{2012AJ....144...39L}, while the objects of study are those $O-C$ values, pulsation parameters, and Fourier coefficients, etc. We introduce the method and process of obtaining these parameters and further analyses in the following subsections.

\subsection{Data Processing and Preanalysis} \label{Sec:Analysis:sub1}

In the $O-C$ analysis of pulsating variable stars, most scholars choose the light maximum as the reference phase, and calculate the times of the corresponding phase. The common method is to select the observation data near the light maximum phase, fit the data with a linear polynomial, and then obtain the times of light maximum by using the calculated first derivatives \citep{2005ASPC..335....3S}. This method is applied to SWASP data: we use the fifth-order polynomial as the fitting formula and determine 25 times of maximum (see Table \ref{Table1}). However, the time resolutions of the photometric data from other sky surveys are lower, so this method is no longer suitable. Therefore, we utilize the same method as in \citet{2014MNRAS.444..600L} to obtain the times of light maximum and minimum. We first divide the light curves into many small segments, and then use the following equation to fit each segment:
\begin{equation}
m(t)=A_{0}+\sum^{n}_{k=1}A_{k}\sin[\frac{2\pi k t}{P_{\rm pul}}+\phi_{k}], \label{Equ:2}
\end{equation}
where $t$ is the time, $A_{0}$ is the mean magnitude, $A_{k}$ is the amplitude of the component $k$, $\phi_{k}$ is its phase in the sine term, and $n$ is the adopted number of terms. Then, according to the fitting results, we can obtain the times of light extrema, pulsation parameters (the full amplitude, $A_{1}$, and magnitude at light maximum $Mag_{\rm max}$), and Fourier coefficients (including the amplitude ratios, $R_{ij}$=$A_{i}/A_{j}$, and the phase differences, $\phi_{ji}=i\phi_{j}-j\phi_{i}$). In the present paper, we provide the values of $R_{21}$, $R_{31}$, $\phi_{21}$, and $\phi_{31}$ of each segment. Because of the different characteristics of sky surveys, the selections of $n$ values are different in specific analyses. For ASAS-SN data, $n$ = 5, and for TESS data, $n$ = 10. For $Kepler$ data, after many attempts, we set $n$ = 10 when determining the times of light extrema to avoid the potential impact of the moir\'{e} effect (see Section \ref{Sec:5}), and set $n$ = 15 when obtaining the Fourier parameters and coefficients (the pulsation amplitude is systematically overestimated when $n$ = 10). This series of analysis methods have been successfully applied \citep{2014MNRAS.444..600L,2018PASJ...70...71L,2021AJ....161..193L,2022MNRAS.510.6050L}.

Table \ref{Table1} lists the times of light maximum of V838 Cyg obtained from the sky surveys. The time standard given by $Kepler$ and TESS is BJD\_TDB. To unify, we convert all the times to HJD\_UTC \citep{2010PASP..122..935E}. The errors of the times of light maximum listed in the second column are obtained based on Method I in \citet{2006Ap&SS.304..363M}. These data are used to analyze the long-term period change of V838 Cyg (see Section \ref{Sec:Analysis:sub2}). To study the weak modulation in V838 Cyg, we also obtained the pulsation parameters and Fourier coefficients from $Kepler$ data, including the times of light maximum and minimum, the full pulsation amplitude, the magnitude at light maximum, $A_{1}$, $R_{21}$, $R_{31}$, $\phi_{21}$, and $\phi_{31}$. These data are provided as supplementary data (see Subsection \ref{Sec:Analysis:sub3}).

\subsection{Long-term period changes} \label{Sec:Analysis:sub2}

\begin{figure}
\centering
\includegraphics[width=.45\textwidth]{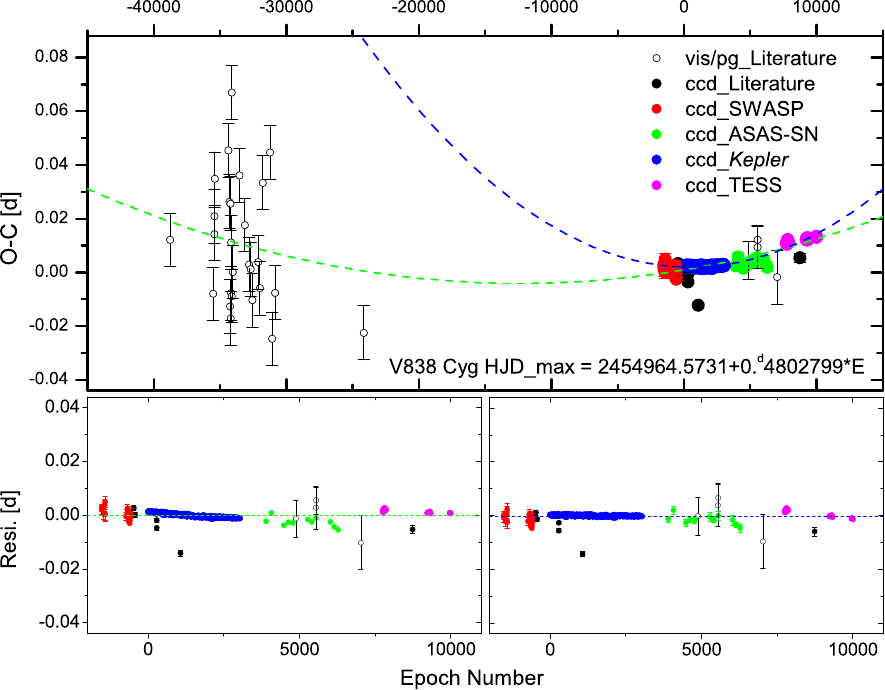}
\caption{The $O-C$ diagram (upper panel) and $O-C$ residuals diagrams (bottom panels) for V838 Cyg. The green and blue dashed lines represent the parabolic variations that reveal the continuous increases in the pulsation period. The former is based on all data, while the latter uses only high-precision CCD data from recent years. The $O-C$ residuals in the bottom left and right panels have removed the parabolic components represented by the green and blue dashed lines respectively. In the bottom left panel, the $O-C$ residuals from $Kepler$ (blue dots) still show variations.} \label{Fig.1}
\end{figure}

Based on the data provided by the GEOS database and $Kepler$, \citet{2011MNRAS.417.1022N} proposed that the pulsation period of V838 Cyg is constant or increases slowly; in the latter case the change in value is $0.05\pm0.04$ day Myr$^{-1}$. Based on the times of light maximum in Table \ref{Table1} and those provided in the literature (\citealt{1970MitVS...5..126M,1978JAVSO...7...79S,2009IBVS.5889....1H,2010IBVS.5941....1H,2011IBVS.5984....1H}, and four visual data provided by J. Vandenbroere and S. Ferrand), we obtained a new $O-C$ diagram (see Figure \ref{Fig.1}) by using the following linear ephemeris of \citet{2011MNRAS.417.1022N}:
\begin{equation}
HJD_{\rm max} = 2454964.5731 + 0.^{\rm d}4802799 \cdot E. \label{Equ:3}
\end{equation}
In Figure \ref{Fig.1}, the $O-C$ diagram shows a clear parabolic shape, which means that the pulsation period increases linearly. Under the two different conditions of whether to consider the early $O-C$ points (Epoch $<$ -20,000), or not, we fitted the $O-C$ diagram with a quadratic polynomial and obtained the corresponding rate of change of period: 0.050(2) day Myr$^{-1}$ (dotted green line) and 0.211(7) day Myr$^{-1}$ (dotted blue line). The former result is consistent with those of \citet{2011MNRAS.417.1022N}, while the latter result without considering the early data is four times larger. It can be seen that although the accuracy of early data is low, it is still very important in the study of long-term period changes. We also plotted the $O-C$ residuals after removing the parabolic component for the two situations (bottom panels in Figure \ref{Fig.1}). It can be seen from the distribution of residuals on the vertical axis that the accuracy of $Kepler$ data is the highest. Therefore, we focus on $Kepler$ data to study the weak modulation components.

\subsection{The modulations in $Kepler$ data} \label{Sec:Analysis:sub3}

\begin{figure*}
\centering
\includegraphics[width=.75\textwidth]{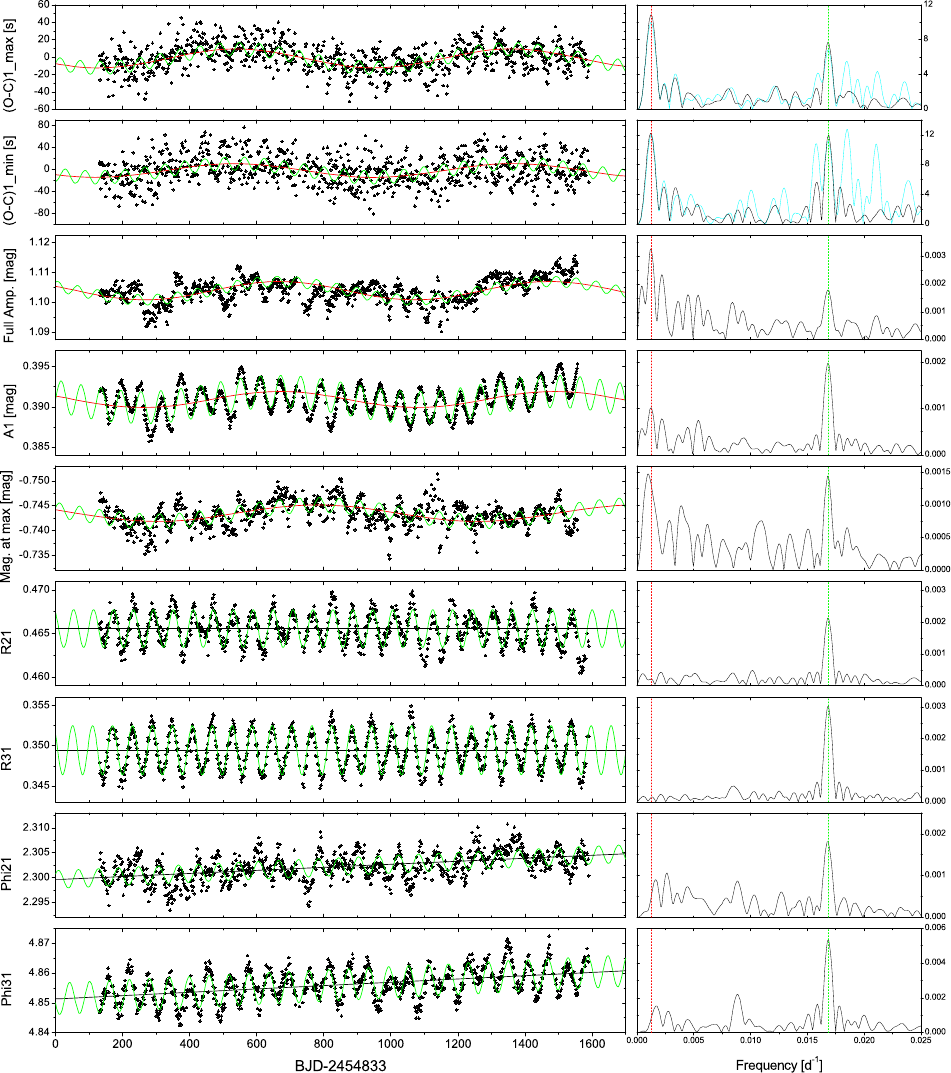}
\caption{Left panels: The modulations of $O-C_{1}$, amplitude, and Fourier coefficient in time series for V838 Cyg. The red solid lines represent the fitting to the long-period modulation, and the green solid lines the fittings to the short-period modulations. Right panels: the corresponding Fourier spectra in the low-frequency range. The red and green dashed vertical lines represent the frequencies of the two modulations. The two cyan spectra in the right upper panels represent the amplitude spectra when $n$ = 15, in which additional peaks appear.} \label{Fig.2}
\end{figure*}

The weak modulation with a period of tens of days in the $Kepler$ light curves of V838 Cyg was discovered by \citet{2013ApJ...773..181N} and \citet{2014ApJS..213...31B}. In this subsection, we use the pulsation parameters and Fourier coefficients to carry out a more detailed analysis. We regard these time series data as signals, process them with a Fourier transform by using the software Period04 \citep{2005CoAst.146...53L}, and obtain the corresponding spectrum information. The panels in the left column of Figure \ref{Fig.2} are the parameters and coefficients plotted as time series, in which $O-C_{\rm 1, max}$ and $O-C_{\rm 1, min}$ denote the $O-C$ residuals after the parabola component has been removed (long-term changes may interfere with the analysis of periodic signals). The red solid lines refer to the change in long-period modulation, while the green solid lines refer to the modulation component with a short period. In the panels of $\phi_{21}$ and $\phi_{31}$, the data also show a linear increase (black solid lines). Therefore, before Fourier transforms, we remove the linear components (3.065$\times$10$^{-6}$ and 5.568$\times$10$^{-6}$ rad day$^{-1}$).

The panels in the right column are the corresponding Fourier amplitude spectra in the low-frequency range. It can be seen that all the spectra show significant peaks at the frequency 0.0168 day$^{-1}$, corresponding to the modulation previously discovered \citep{2014ApJS..213...31B}. However, in the upper five spectra, it is obvious that there is also a peak with lower frequency (the mean value is about $0.00119\pm0.00003$ day$^{-1}$, and the corresponding period is about $840\pm21$ days), but there is no corresponding component in the amplitude spectra of $R_{21}$, $R_{31}$, $\phi_{21}$, and $\phi_{31}$. This suggests that the low-frequency component seems to be some simpler modulation. The red and green vertical dotted lines indicate the frequencies of the two modulation components. In the upper right two panels, the cyan line represents the amplitude spectra when $n$ = 15 (see the introduction in Subsection \ref{Sec:Analysis:sub1}). It can be seen that there are several peaks around the short-period modulation, and their frequencies are 0.01577, 0.01684, 0.01848, and 0.02105 day$^{-1}$ (the corresponding periods are 63, 59, 54, and 47 days), respectively. This means that, when $n$ = 15, the analysis results are affected by the moir\'{e} effect (see Section~\ref{Sec:5} for details).

\begin{figure*}
\centering
\includegraphics[width=.35\textwidth]{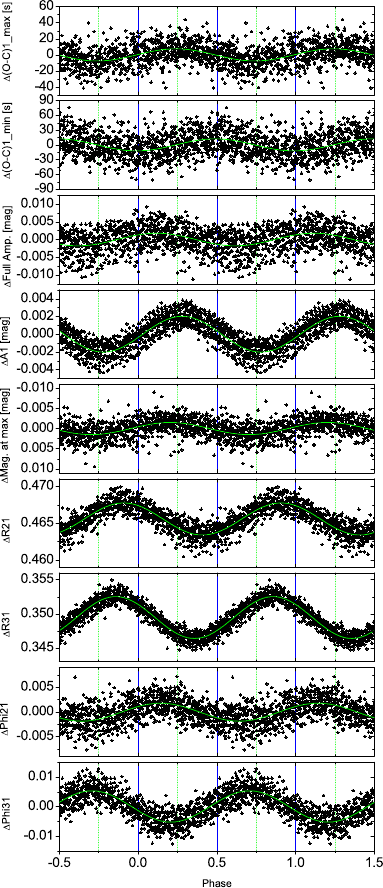}
\caption{Phase diagrams for the short-period modulation. The phase difference between $O-C_{\rm 1, max}$ and $O-C_{\rm 1, min}$ is about $\pi/2$, while the changes between $O-C_{\rm 1, max}$ and $A_{1}$ (and $Mag_{\rm max}$) are basically in phase. In addition, $R_{21}$ is in phase with $R_{31}$, while $\phi_{21}$ and $\phi_{31}$ are just opposite.} \label{Fig.3}
\end{figure*}

\begin{figure*}
\centering
\includegraphics[width=.35\textwidth]{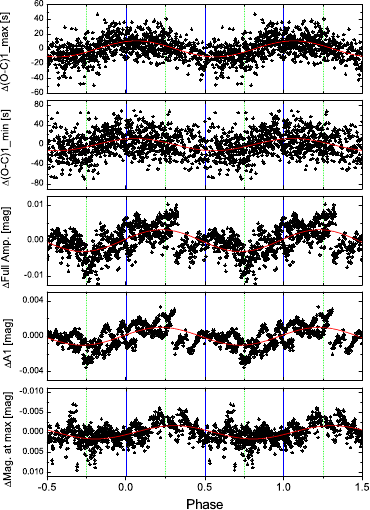}
\caption{Phase diagrams for the long-period modulation. The phases of $O-C_{\rm 1, max}$ and $O-C_{\rm 1, min}$ are the same, and the changes in the three parameters related to the amplitude, full amplitude, $A_{1}$, and magnitude at light maximum are basically in phase.} \label{Fig.4}
\end{figure*}

To further study the modulations, we also plot their parameter phase diagrams in Figure \ref{Fig.3} and \ref{Fig.4}. Figure \ref{Fig.3} shows the phase diagrams of short-period modulation. It can be seen that the phase difference between $O-C_{\rm 1, max}$ and $O-C_{\rm 1, min}$ is about $\pi/2$, while the changes between $O-C_{\rm 1, max}$ and $A_{1}$ (and $Mag_{\rm max}$) are basically in phase. In addition, we also found that the $R_{21}$ is in phase with $R_{31}$, while $\phi_{21}$ and $\phi_{31}$ are just opposite. For the long-period modulation (see Figure \ref{Fig.4}), the phases of $O-C_{\rm 1, max}$ and $O-C_{\rm 1, min}$ are the same, and the changes in the other three parameters related to the amplitude, full amplitude, $A_{1}$, and magnitude at light maximum are basically in phase. Table \ref{Table2} lists the corresponding parameter results and their estimated errors. The parameters are obtained by fitting with double sine functions. In actual analysis, the software Period04 was used, and the method for calculating the errors of frequencies was provided by \citet{1999DSSN...13...28M}. The mean period values calculated from the mean frequencies are listed in the bottom row of Table~\ref{Table2}. An error propagation formula was used to calculate the errors of the mean period.

\begin{figure}
\centering
\includegraphics[width=.45\textwidth]{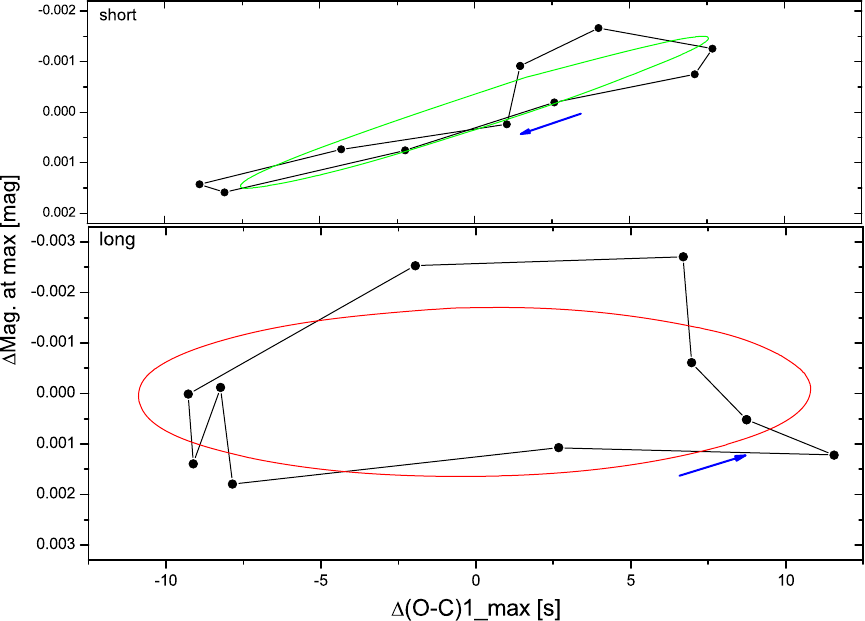}
\caption{Merged magnitude at light maximum (ordinate) vs. the corresponding $O-C_{\rm 1, max}$ values (abscissa) for the two modulations. The solid green and red lines denote the fitting results (see Table \ref{Table2}). For the short-period modulation (upper panel), the closed curve runs in a clockwise direction, while it runs counterclockwise for the long-period modulation (bottom panel).} \label{Fig.5}
\end{figure}

To demonstrate the varieties of phenomenology, \citet{2012AJ....144...39L} used the $O-C$ and magnitude at light maximum folded with modulation period to plot the closed curves that describe the shape of the Blazhko effect. Using their method, we also plot the $O-C_{\rm 1, max}$ versus $Mag_{\rm max}$ diagram of the two modulations of V838 Cyg in Figure \ref{Fig.5}. The black points refer to the binned observation points, while the green and red solid lines are the corresponding fitting results. For the short-period modulation, the change is clockwise, and the two parameters are roughly in phase, which is consistent with the situation in Fig. 9 of \citet{2012AJ....144...39L}. The change in long-period modulation is counterclockwise, and the correlation between $O-C_{\rm 1, max}$ and $Mag_{\rm max}$ is not obvious, showing that it is indeed different from the short-period modulation.

\section{The modulations in V838 Cyg}  \label{Sec:4}

The light curves of V838 Cyg show two different weak modulations. The component with a short period causes the changes in pulsation period, amplitude, and the Fourier coefficients, while the component with a long period only causes the change in period and amplitude. The former exhibits typical Blazhko modulation characteristics, and it seems that it is intrinsic (due to the changes in the stellar atmosphere, i.e., multimode resonance, \citealt{2010MNRAS.409.1244S,2011ApJ...731...24B,2011MNRAS.414.1111K,2015ApJ...802...52B}; the variation of turbulent convection strength due to oscillatory magnetic field, \citealt{2006ApJ...652..643S,2010PASP..122..536S}; or shock waves, \citealt{2013A&A...554A..46G}), while the latter may be extrinsic (such as due to the light travel time effect, LiTE, \citealt{1952ApJ...116..211I}).

\begin{figure*}
\centering
\includegraphics[width=.75\textwidth]{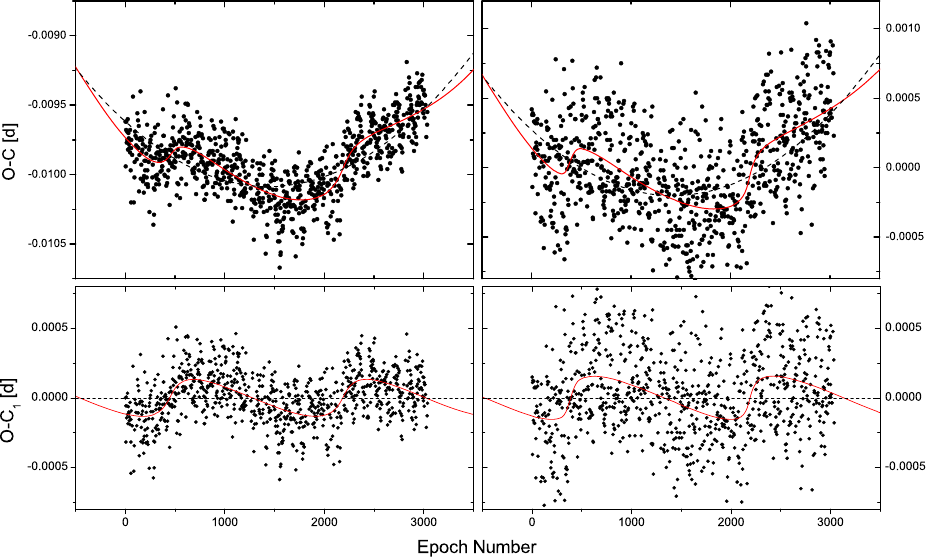}
\caption{Upper panels: $O-C$ diagram of V838 Cyg for the pulsation maxima (left) and minima (right). The red solid lines refer to a combination of upward parabolic and periodic variations. The black dashed lines represent the parabolic variation that reveals a continuous increase in the pulsation period. Bottom panels: the $O-C$ residuals from the quadratic term to V838 Cyg for the pulsation maxima (left) and minima (right). The red solid lines referring to the periodic variation can be seen clearly.} \label{Fig.6}
\end{figure*}

\citet{2014MNRAS.444..600L} found similar changes in the $O-C$ diagrams of the other two RR Lyrae stars FN Lyr and V894 Cyg in the $Kepler$ field. According to the same trends in changes in the $O-C$ diagrams obtained from the times of light maximum and minimum, they supposed the mechanism of the changes is the LiTE caused by the presence of a companion star. We assume that this effect also exists in V838 Cyg, use an eccentric orbit model to fit the $O-C$ diagrams \citep{2014MNRAS.444..600L}, and obtain the partial orbital parameters (see Table \ref{Table3}). Figure \ref{Fig.6} plots the corresponding $O-C$ diagrams, in which the same trends in changes can be observed.

In this case, assuming the mass of the pulsating primary star is 0.6 M$_{\bigodot}$, the calculated lower mass limit of the companion is about 0.0126 M$_{\bigodot}$, and it should be a massive planet. But one problem that needs to be considered is whether the presence of a planet causes amplitude variation and a change in $Mag_{\rm max}$. The general viewpoint is that a companion with a long orbital period will only cause modulation of the pulsation frequency \citep{2012MNRAS.422..738S}. Eclipsing may occur, but it is difficult to recognize because of the interference of pulsation. However, in the field of binary stars and cataclysmic variable stars, there is a reflection effect between companions \citep{1990ApJ...356..613W}, and perhaps the companion can influence the pulsation amplitude and $Mag_{\rm max}$ of the pulsating host star through this effect.

The long-term weak $O-C$ changes found in the $Kepler$ field were believed to be caused by an instrumental effect \citep{2019MNRAS.485.5897B}. The most important factor should be the $Kepler$ frequency $f_{\rm K}$ (it is the reciprocal of the $Kepler$-year, \citealt{2013MNRAS.436.1576B}). This phenomenon is manifested in the $Kepler$ light curves of some stars that show variability with a period equal a $Kepler$-year, and the amplitude and phase of the variations are related to the position of the targets in the $Kepler$ field of view (see Figure 5 and 6 of \citealt{2013MNRAS.436.1576B}). This means that the $Kepler$-year mainly manifests in its impact  on the flux (or magnitude), which may interfere with the related parameters (such as $Mag_{\rm max}$ in this paper). The variations of $O-C$ values mainly reflect the changes in pulsation frequency/period \citep{2011MNRAS.417..974B}, and the effect of amplitude modulation is small. Moreover, in our analysis, the long-cadence data were detrended, and the factor of a $Kepler$-year can be effectively removed before the $O-C$ analysis.

\begin{figure}
\centering
\includegraphics[width=.45\textwidth]{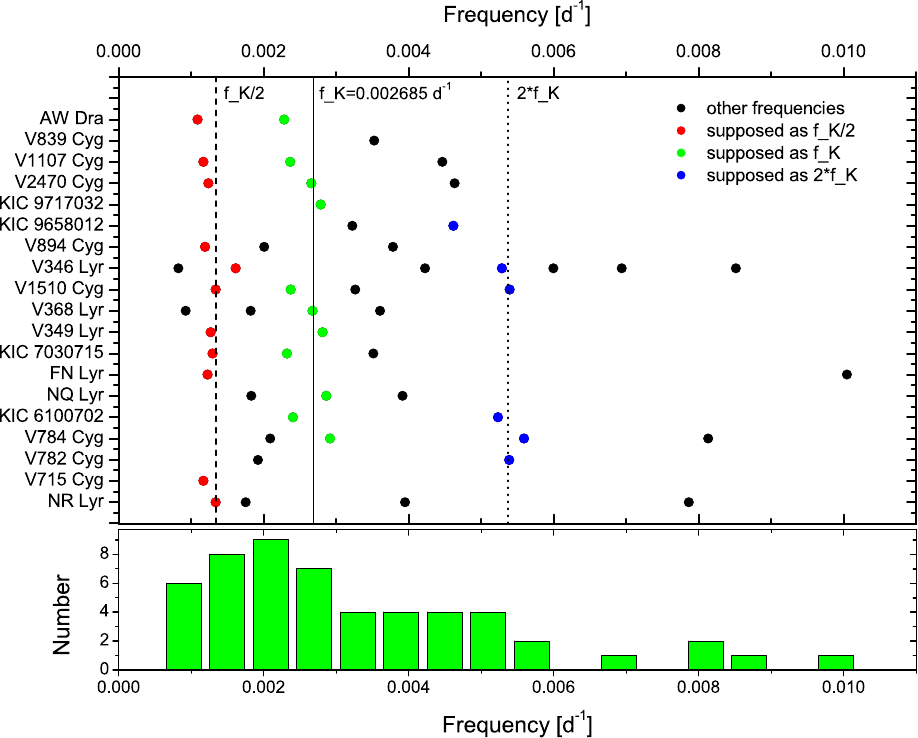}
\caption{The frequency distribution of the components listed in Table 4 of \citet{2019MNRAS.485.5897B}. The red dots in the upper panel indicate those components belong that to $f_{\rm K}/2$, green dots those to $f_{\rm K}$, blue dots those to $2f_{\rm K}$, and the black dots represent other detected components. The bottom panel shows the histogram of all components.} \label{Fig.7}
\end{figure}

\citet{2019MNRAS.485.5897B} pointed out that the components in the low-frequency spectral region of the $O-C$ diagrams for the non-Blazhko RRab stars in the $Kepler$ field are mostly harmonics or subharmonics of the $Kepler$ frequency. Therefore, it can be expected that the frequencies in the corresponding spectral diagrams should be strictly equal to the $Kepler$ frequency and the harmonics. However, their $O-C$ diagrams are different, and the corresponding frequency values are also different (see Figure 13 and Table 4 in \citealt{2019MNRAS.485.5897B}). We conducted statistical analysis on the frequency components given in Table 4 of \citet{2019MNRAS.485.5897B}, and plotted Figure~\ref{Fig.7}. The red dots in the upper panel indicate those components that belong to $f_{\rm K}/2$, green dots those to $f_{\rm K}$, blue dots those to $2f_{\rm K}$, and the black dots represent other detected components. The bottom panel shows the histogram of all components. It can be seen that the components within the range 0 - 0.006 day$^{-1}$ are almost uniformly distributed. It is unnecessary to consider those components with frequencies close to $f_{\rm K}/2$, $f_{\rm K}$, and $2f_{\rm K}$ as related to instrumental factors.

In fact, in other fields of variable star research, scholars have found weak variations in the $O-C$ diagrams based on the $Kepler$ data and conducted further and deeper research \citep{2013ApJ...774...81T,2013MNRAS.432.2284M,2012MNRAS.425.1312D,2017MNRAS.464.1553D}. The $Kepler$ photometry data are still some of the most accurate data that we can obtain. Researchers should be encouraged to dig deeper, analyze, and research them. Based on our analyses and discussions, the long-period modulation in V838 Cyg shows different characteristics from the general Blazhko modulation. We initially propose that the corresponding mechanism is the LiTE; at least it should be an external factor.

\section{Moir\'{e} patterns and beat frequencies}  \label{Sec:5}

Due to the characteristics of $Kepler$ data, the long-cadence light curves of many variable stars show a significant moir\'{e} pattern. We found that the moir\'{e} effect may have a greater impact on the corresponding analysis than the $Kepler$-year. We will provide a detailed introduction to this effect in this section and explore its interference in V838 Cyg analysis.

The $Kepler$ observations are made in two cadences: long cadence (LC) and short cadence (SC). The LC data, which result from 270 exposures coadded with a total integration time of 29.4 min, have a low time resolution, but are given continuously. For SC data, nine frames coadded with a total of 58.85 s are given in some Quarters. The original data available at the Multimission Archive at the Space Telescope Science Institute (STScI)\footnote{\url{http://archive.stsci.edu/kepler/}} consist of raw and corrected flux counts as a function of time. Corrected fluxes have been processed by the presearch data conditioning (PDC) pipeline, and the signatures in the light curves that are correlated with systematic error sources have been removed\footnote{\url{https://archive.stsci.edu/kepler/manuals/Data\_Characteristics.pdf}}. \citet{2014ApJS..213...31B} and \citet{2015ApJ...809L..19B} used the tailor-made apertures to process the $Kepler$ pixel photometric data, and presented rectified light curves of some RR Lyrae stars in the $Kepler$ field\footnote{\url{http://www.konkoly.hu/KIK/data\_en.html}}. In this analysis, we use corrected PDC fluxes mainly, but use the rectified data of V838 Cyg for more accurate results. The data processing and reduction methods used here are the same as those described in \citet{2014MNRAS.444..600L}. We use the linear ephemerides in Table~\ref{Table1App} (columns 2 and 3) to calculate the $O-C$ values of stars.

\begin{figure*}
\includegraphics[width=17cm]{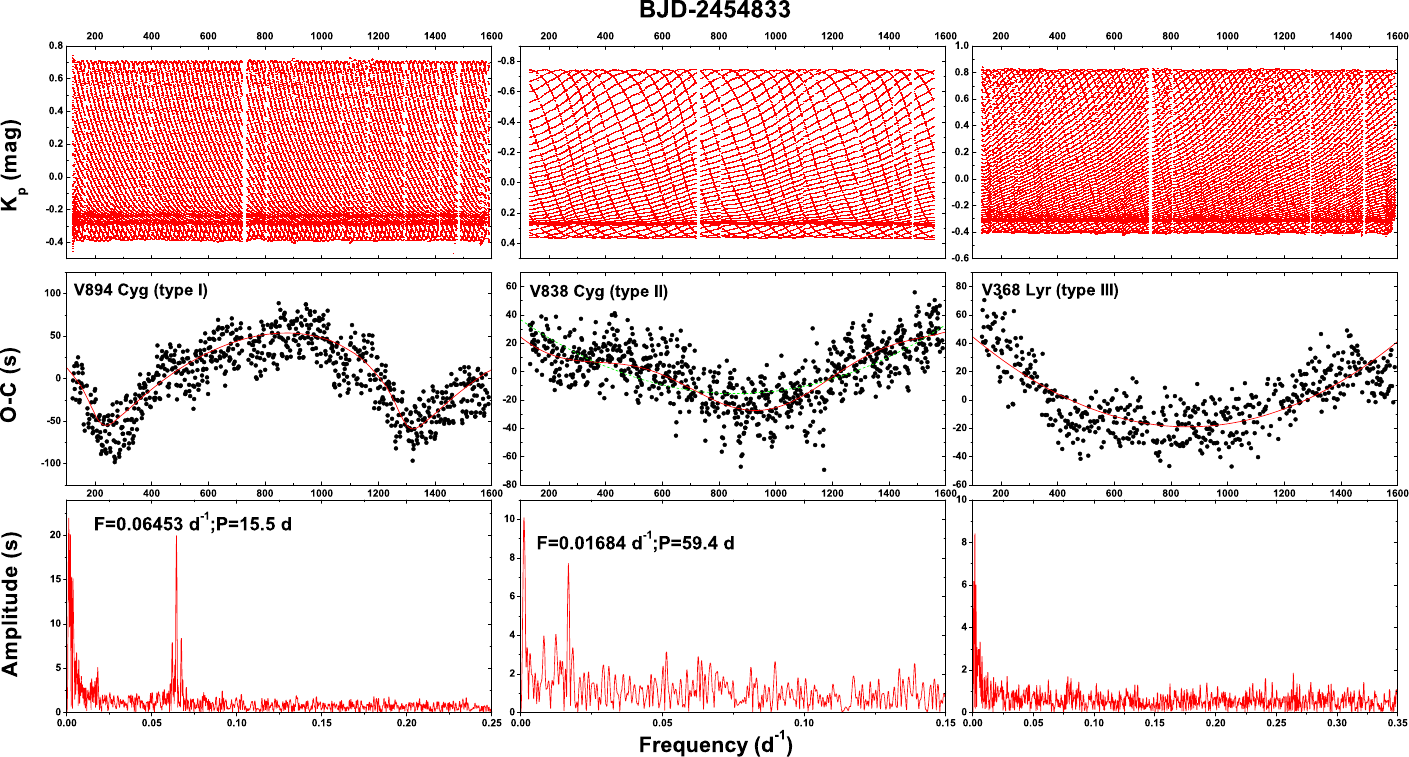}
\caption{Top panels: Light curves of three RRab stars in the $Kepler$ field that show clear moir\'{e} patterns. Middle panels: The corresponding $O-C$ diagrams. The red lines represent the fittings to the $O-C$ diagrams. Bottom panels: The Fourier spectra of the $O-C$ diagrams. The prominent components are labeled. But no peak is found in the spectrum of V368 Lyr.} \label{Fig.8}
\end{figure*}

The moir\'{e} patterns exhibited in the $Kepler$ LC data of RR Lyrae stars are common. One of the conditions to produce these phenomena is that the ratio of the pulsation period of the RR Lyrae star ($P_{\rm pul}$) and the LC sampling period ($\Delta t_{\rm LC} \simeq$ 1765.5 s $\simeq$ 29.4 minutes) is close to an integer (e.g. for V894 Cyg, $P_{\rm pul}/\Delta t_{\rm LC}$ = 27.96309 $\simeq$ 28, \citealt{2011MNRAS.417.1022N}). Column 4 in Table~\ref{Table1App} shows the examples. These moir\'{e} patterns have a remarkable influence on the $O-C$ analysis: prominent beat components are observed in the $O-C$ curves and corresponding Fourier spectrum \citep{2013ApJ...768...33R}. As an example, the left panels in Figure~\ref{Fig.8} show the light curves and the $O-C$ diagram of V894 Cyg. The moir\'{e} pattern can be seen clearly in the light curve.  In addition, there is spurious periodicity exhibited in the $O-C$ curves. This periodicity cannot be observed clearly, but causes the $O-C$ points to show a diffuse distribution in the vertical direction. For the sake of a better view, we present the Fourier spectrum of the $O-C$ diagram (see the bottom left panel in Figure~\ref{Fig.8}). One can observe the significant peak in the spectrum. Column 5 of Table~\ref{Table1App} lists the corresponding frequencies. Moreover, in the $O-C$ diagram, the $O-C$ curves show clear periodic variations on a longer timescale, which has been explained as a consequence of the LiTE caused by the star following an elliptical orbit in a binary system \citep{2014MNRAS.444..600L}.

\citet{2013ApJ...768...33R} gave the formulae to calculate this type of beat frequencies:
\begin{equation}
f_{\rm b1,1}=f_{\rm LC}-f_{\rm pul} \cdot \rm
int(\frac{\emph{f}_{LC}}{\emph{f}_{pul}}), \label{Equ:4}
\end{equation}
or
\begin{equation}
f_{\rm b1,2}=f_{\rm pul}-f_{\rm b1,1}, \label{Equ:5}
\end{equation}
where ``int" gives the values, which are rounded down to the nearest integer, $f_{\rm pul}$ is the pulsation frequency, and $f_{\rm LC}
\equiv 1/\Delta t_{\rm LC}$. Column 6 of Table~\ref{Table1App} contains the two beat frequencies, and those calculated values labeled with asterisks are consistent with the results given by the Fourier analysis (column 5 of Table~\ref{Table1App}).

All of these stars match this condition except two: V838 Cyg and V368 Lyr. Their light curves both clearly show moir\'{e} patterns (see the top middle panel and top right panel in Figure~\ref{Fig.8}), but their ratios ($P_{\rm pul}/\Delta t_{\rm LC}$) are close to half-integer (23.50442) and 1/3-integer (22.33997), respectively. According to this feature, we divide the moir\'{e} patterns into three types (type I for integer, type II for half-integer, and type III for 1/3-integer; we also investigate those ratios close to 1/4-integer (e.g. NQ Lyr, KIC 9717032), but the scalloped patterns at extremum are no longer clear; so only these three types are discussed in this section).

\begin{figure*}
\centering
\includegraphics[width=8.5cm]{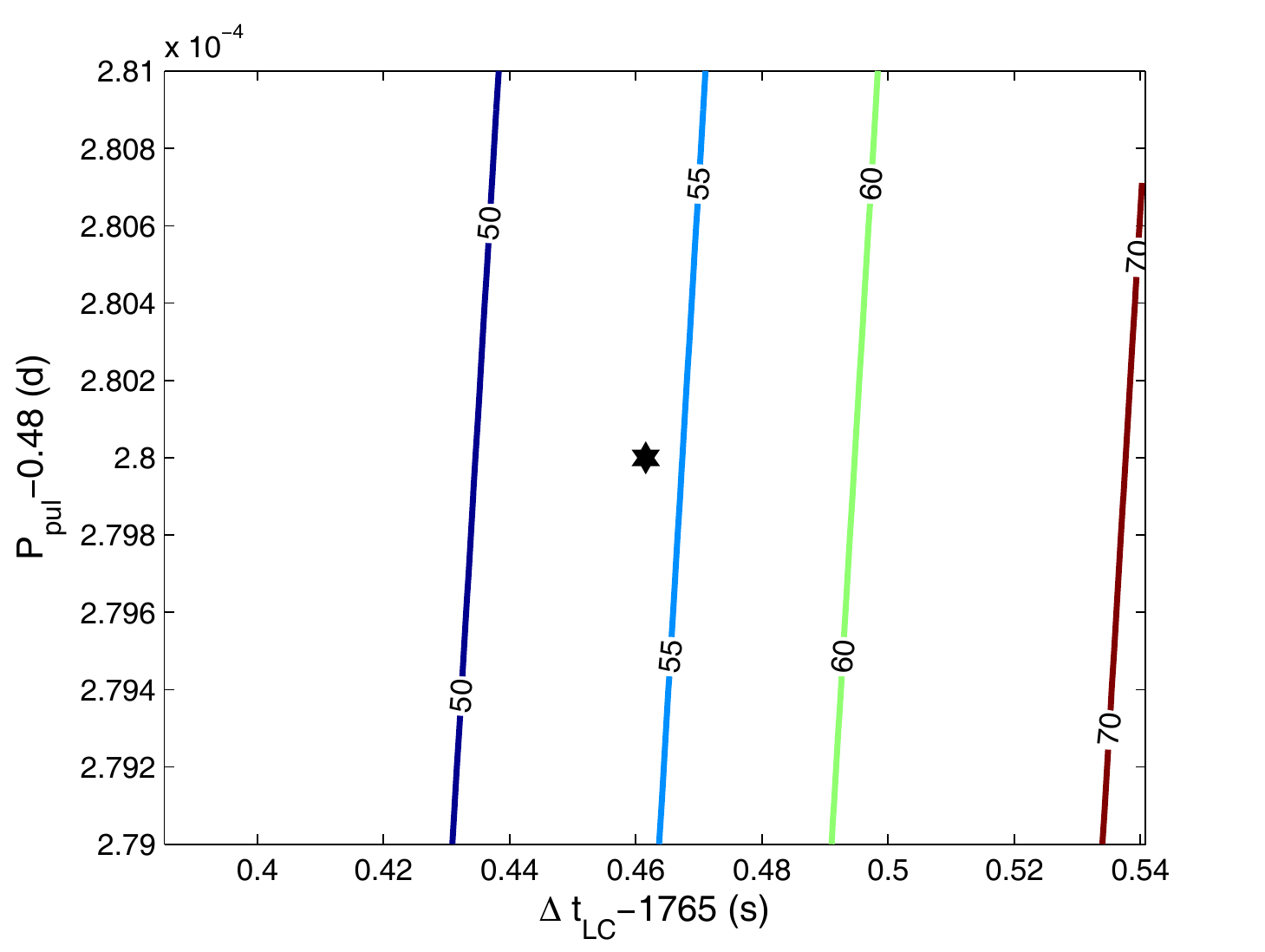}
\caption{Contour map of the beat period. The asterisk denotes the mean value of beat period (54.3 days), where $\Delta t_{\rm LC}=1765.46$ s and $P_{\rm pul}=0.48028$ days. It can be seen that the beat period $P_{\rm beat}$ is more sensitive to $\Delta t_{\rm LC}$ than to $P_{\rm pul}$.} \label{Fig.9}
\end{figure*}

\begin{figure*}
\centering
\includegraphics[width=8.5cm]{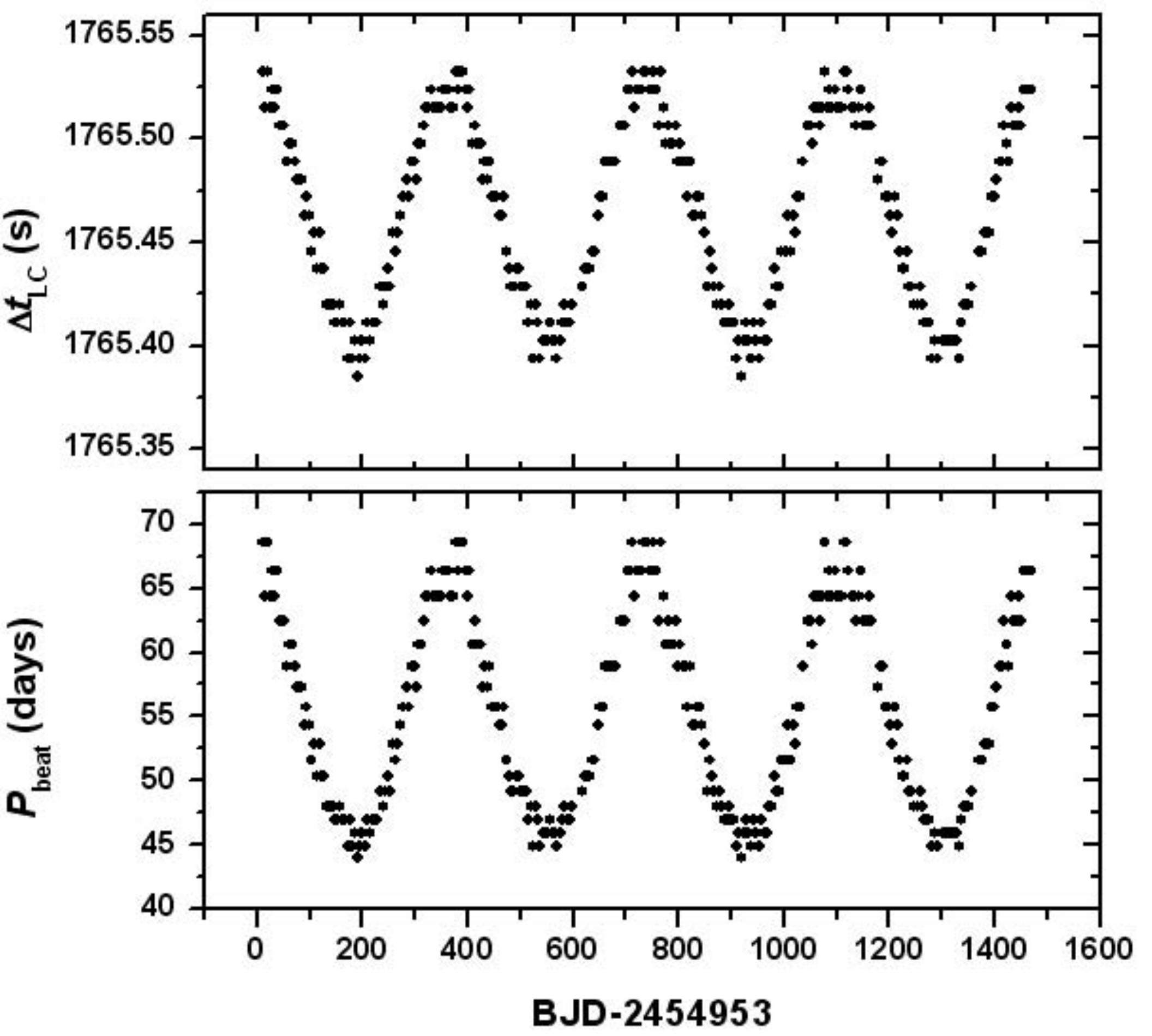}
\caption{$\Delta t_{\rm LC}$ (top panel) and beat period $P_{\rm beat}$ (bottom panel) with time. The values of $\Delta t_{\rm LC}$ are obtained by data time subtracting its previous data time that given in LC data. In the calculation, the pulsation period $P_{\rm pul}$ is set as constant, 0.48028 days. The values of $P_{\rm beat}$ are obtained by incorporating $\Delta t_{\rm LC}$ and $P_{\rm pul}$ into Equations (\ref{Equ:4}, \ref{Equ:5}, \ref{Equ:6}, \ref{Equ:7}). It can be seen that the values of $P_{\rm beat}$ range from 45 to 66 days.} \label{Fig.10}
\end{figure*}

The beat frequencies of V838 Cyg (type II) and V368 Lyr (type III) can be calculated as:
\begin{equation}
f_{\rm b2}=|f_{\rm b1,1}-f_{\rm b1,2}| \label{Equ:6}
\end{equation}
and
\begin{equation}
f_{\rm b3}=|2 f_{\rm b1,1}-f_{\rm b1,2}|. \label{Equ:7}
\end{equation}
Column 6 of Table~\ref{Table1App} lists the beat frequencies. From Equations (\ref{Equ:6}, \ref{Equ:7}), type II and III beat frequencies can be described as the further beats of type I beat frequencies, and their intensities can be expected to be weaker than type I. The beat frequency of V838 Cyg is 0.0184 day$^{-1}$, and the beat period is about 54.3 days. This result agrees with \citet{2013ApJ...773..181N}. However, \citet{2013ApJ...773..181N} also pointed out that the modulation period for V838 Cyg was varying, and they identified multiple periods of 54, 64, and 47 days. The reason for this phenomenon should be the small changes of $\Delta t_{\rm LC}$ with time. From Equations (\ref{Equ:4}, \ref{Equ:5}, \ref{Equ:6}), we can see that the beat frequency in V838 Cyg is dependent on $f_{\rm pul}$ and $f_{\rm LC}$ (i.e. $P_{\rm pul}$ and $\Delta t_{\rm LC}$), both of which may change with time. In order to study the relationship between the two parameters and the beat period, we provide the corresponding contour map in Figure~\ref{Fig.9}. It can be seen that the contour lines are basically vertical, which means that the gradient direction is basically horizontal. This indicates that when the $\Delta t_{\rm LC}$ on the abscissa changes while $P_{\rm pul}$ on the ordinate remains unchanged, the beat period changes the most, that is, the beat period is more sensitive to $\Delta t_{\rm LC}$ than to $P_{\rm pul}$. Therefore, in the following discussion, we set $P_{\rm pul}$ as a constant (= 0.48028 days), and only explore the impact of changes in $\Delta t_{\rm LC}$ on the beat period. Actually, the $\Delta t_{\rm LC}$ changes periodically in one $Kepler$-year ($P_{K}=372.5$ days), ranging from 1765.38528 to 1765.54080 s (see the top panel in Figure~\ref{Fig.10}, these values are obtained by data time subtracting its previous data time that listed in LC data). We substitute these values into Equations (\ref{Equ:4}, \ref{Equ:5}, and \ref{Equ:6}), and obtain the values of the beat period with time (see the bottom panel in Figure~\ref{Fig.10}). We fit these values with the following function:
\begin{equation}
P_{\rm beat}(t)=P_{\rm b0}+A_{\rm b0} \sin(\frac{2 \pi}{P}t+T_{\rm b0}), \label{Equ:8}
\end{equation}
and the results are: $P_{\rm b0}=55.34$ days, $A_{\rm b0}=10.41$ days, $P=372.28$ days ($\simeq$ one $Kepler$-year), and $T_{\rm b0} = -0.3964$. So the beat periods range from 45 to 66 days and the mean value is 55 days, which agrees with the dominant values of modulation period found by \citet{2013ApJ...773..181N}. Moreover, \citet{2013ApJ...773..181N} pointed out that the waves are compressed around times $t=200$ and 550 (BJD 2,455,153 and 2,555,503) but were wider around $t=400$ and 750 (BJD 2,455,353 and 2,455,703). Comparing these values with the curve in Figure~\ref{Fig.10}, they also agree with each other. So it is safe to say that the results of V838 Cyg given were influenced by the moir\'{e} pattern in the LC light curves.

In the Fourier spectrum of the $O-C$ diagram calculated from LC data, we find a frequency of $f_{\rm beat}=0.01684\pm0.00003$ day$^{-1}$, or $P_{\rm beat}=59.4\pm0.1$ days (see middle panels in Figure~\ref{Fig.8}). This agrees with the result obtained by \citet{2014ApJS..213...31B}. However, different from those type I beat frequencies, the frequency in the Fourier spectrum of V838 Cyg is not virtual, but is real modulation. It is a complete coincidence that the beat frequency (0.0184 day$^{-1}$) and the modulation frequency (0.01684 day$^{-1}$) are close to each other. 

The beat frequency of V368 Lyr is 0.0436 day$^{-1}$. But in the Fourier spectrum, we do not find a significant peak at the corresponding position (see the bottom right panel in Figure~\ref{Fig.8}). However, we can obtain the period by counting the patterns in light curves, and the result is about 24 days, which roughly agrees with the calculated beat period (22.9 days). From our work, the influences of type II and III beat frequencies on the $O-C$ analysis are much weaker than those of type I. In most cases, they can be ignored safely, but sometimes they may influence the $O-C$ analysis, just like the situation in V838 Cyg.

\begin{figure*}
\centering
\includegraphics[width=.9\textwidth]{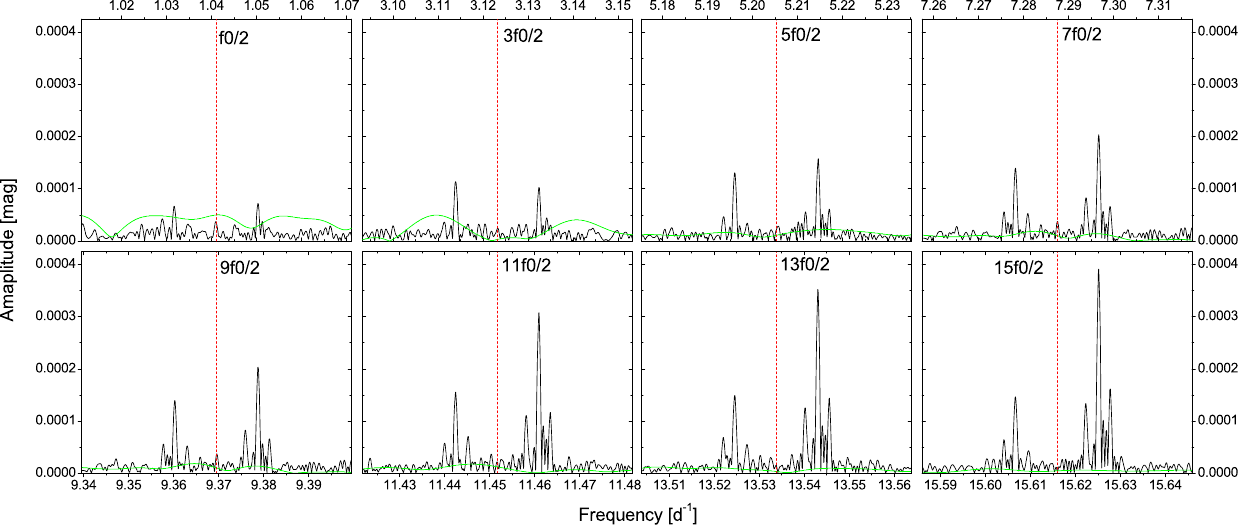}
\caption{Fourier spectra of V838 Cyg around the half-integer pulsation frequencies after the light curve data were pre-whitened with the main pulsation frequencies. Black and green lines denote the spectra from $Kepler$ LC and SC data, respectively. The vertical red dashed lines show the positions of exact half-integer frequencies. The double peaks in each panel can be identified as moir\'{e} side peaks: ($f_{\rm b1,1}$, $f_{\rm b1,2}$), ($f_{0}+f_{\rm b1,1}$, $f_{0}+f_{\rm b1,2}$), ($2f_{0}+f_{\rm b1,1}$, $2f_{0}+f_{\rm b1,2}$), etc.} \label{Fig.11}
\end{figure*}

In addition, in the case of V838 Cyg, the moir\'{e} effect will also have an impact on the study of the period doubling phenomenon. We performed Fourier analysis on both LC and SC $Kepler$ data of V838 Cyg. After pre-whitening the data with the pulsation frequency and its harmonics, the amplitude spectra near the half-integer pulsation frequencies ($0.5f_{0}$, $1.5f_{0}$, ...) are presented in Figure~\ref{Fig.11}. The black spectrum represents the result of LC data, and the green line represents the spectrum of SC data. It can be seen that there is no peak on the half-integer pulsation frequencies. But for LC data, there are double peaks around the half-integer frequencies with a frequency difference of 0.0184 day$^{-1}$. These components are caused by the moir\'{e} effect, but are easy to consider as half-integer pulsation frequencies. The light curve of V838 Cyg does not actually show the period doubling phenomenon, and the moir\'{e} pattern misled the previous study \citep{2014ApJS..213...31B}.

\section{Metal abundance and Blazhko modulation} \label{Sec:6}

\begin{figure}
\centering
\includegraphics[width=.45\textwidth]{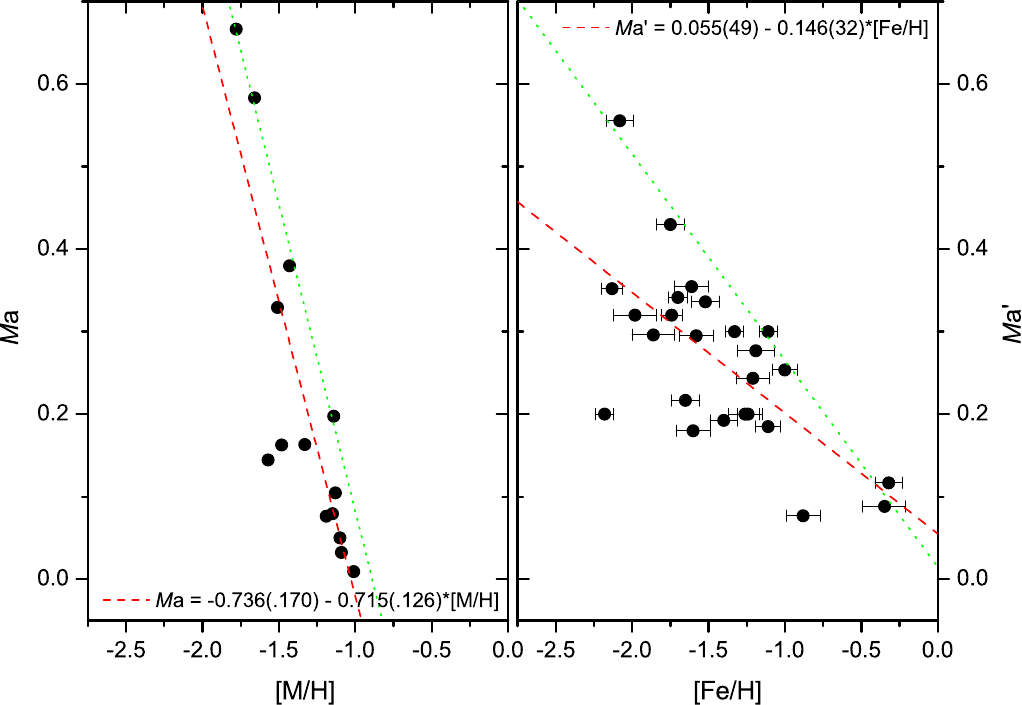}
\caption{Metal abundance vs. amplitude modulation factor. The red dashed lines refer to the linear fitting to the data, and the green dotted lines represent the arbitrary upper limit on the modulation factor. [M/H] and [Fe/H] are obtained from \citet{2013ApJ...773..181N} and \citet{1994AJ....108.1016L}, and $M_{\rm a}$ and $M^{'}_{\rm a}$ are calculated based on the relevant pulsation amplitudes provided by \citet{2013ApJ...773..181N} and \citet{2012AJ....144...39L}. Only \citet{1994AJ....108.1016L} provided the errors of [Fe/H] for each stars (see the horizontal error bars in the right panel).} \label{Fig.12}
\end{figure}

One of our aims is to explore the relationship between metal abundance and Blazhko modulation. Noting that V838 Cyg is the RRab star with the highest metal abundance and the weakest modulation among all Blazhko RRab stars in the $Kepler$ field \citep{2013ApJ...773..181N}, we define a similar parameter with reference to the amplitude modulation factor (characteristic modulation depth) in radio technology:
\begin{equation}
M_{\rm a} = \frac{A_{\rm full, max}-A_{\rm full, min}}{A_{\rm full, max}+A_{\rm full, min}}, \label{Equ:9}
\end{equation}
where $A_{\rm full, max}$, and $A_{\rm full, min}$ are the maximum and minimum values of the full amplitude of the light curve, respectively. These values and the metal abundances are given in Tables 2 and 9 of \citet{2013ApJ...773..181N}. In the left panel of Figure \ref{Fig.12} the metal abundances are plotted against $M_{\rm a}$. It can be seen that there is a significant negative correlation between the two parameters, that is, the higher the metal abundance, the weaker the Blazhko modulation. The red dashed line in the left panel refers to a linear fitting to the data, and the corresponding result is
\begin{equation}
M_{\rm a} = -0.736(.170)-0.715(.126)[M/H]. \label{Equ:10}
\end{equation}
The correlation coefficient of the model is -0.85. We also notice that all those RRab stars in the $Kepler$ field with higher metal abundance than V838 Cyg are non-Blazhko stars, and there are also some non-Blazhko stars that show low metal abundances. These phenomena seem to indicate that metal abundance does not necessarily determine whether the modulation occurs, but when it occurs, the metal abundance has an important influence on the intensity of modulation: rich metals will 'suppress' modulation.

The sample size of $Kepler$ RR Lyrae stars is not large, and more information is collected in other literature. \citet{2012AJ....144...39L} provided the magnitude variation at the light maximum of some Blazhko stars (see their Table 1). We define a similar parameter:
\begin{equation}
M^{'}_{\rm a} = \frac{\Delta Mag_{\rm max}}{A_{\rm full, mean}}, \label{Equ:11}
\end{equation}
The right panel of Figure \ref{Fig.12} plots the metal abundance against the $M^{'}_{\rm a}$, where the metal abundances of the corresponding stars are given by \citet{1994AJ....108.1016L}. A similar trend can be seen in the diagram. The red dashed line in the right panel refers to a linear fitting, and the corresponding result is
\begin{equation}
M^{'}_{\rm a} = -0.055(49)-0.146(32)[Fe/H]. \label{Equ:12}
\end{equation}
The correlation coefficient of the model is -0.68. However, unlike the left panel, there are also some stars with weak modulations located in the range of poor metal abundance. But in the metal-rich range, the modulations are the weakest. This illustrates the 'suppression' effect: large modulation is restricted to high metallicity.

The next question is how the metal abundance affects the Blazhko modulation. We are not familiar with the relevant theories and models, but only give a superficial viewpoint: in the stellar evolution model, the metal abundance is generally considered to mainly affect the opacity: greater metal abundance makes the opacity increase, the radiation temperature gradient increases accordingly, and this affects the convection in the atmosphere (intensity and boundary position). In this case, the turbulence, magnetic field activity, and shock wave that may cause the Blazhko modulation are suppressed. Based on observation, \citet{2014AJ....148...88C} proposed that the Blazhko effect originates from a dynamical interaction between a multishock structure and an outflowing wind in a coronal structure; and then \citet{2017ApJ...835..187C} showed that the main shock intensity of metal-poor RR Lyrae stars is larger than that of metal-rich stars. According to their viewpoints, there is the following logical relationship: the metal abundance in the stellar atmosphere is negatively correlated with the shock intensity, thus suppressing the modulation caused by the shock waves.

The metal abundance is related to several other stellar intrinsic parameters (such as luminosity, mass, and effective temperature) and pulsation parameters. Whether some of these parameters are more closely related to modulation needs to be considered. We also compared the relationship between $M_{\rm a}$ ($M^{'}_{\rm a}$) and other pulsation parameters ($P_{\rm pul}$, $P_{\rm Bl}$, and other stellar atmospheric parameters), but did not find any significant correlation as shown in Figure \ref{Fig.12}. However, this cannot be ruled out because some physical quantities are also interfered with by the effect of evolution. \citet{2022MNRAS.510.6050L} studied the four RRc stars in $Kepler$ field, and found that the intensities of changes in the $O-C$ diagram are related to the macroturbulent velocities. In fact, the metal abundances of those four stars are also negatively correlated with the macroturbulent velocity.

RR Lyrae stars with low metal abundance have greater luminosities and higher masses \citep{2007A&A...476..779B,2010ApJ...722...79S,2013ApJ...773..181N}. If the correlations in Figure \ref{Fig.12} are established, it would indicate that strong modulation tends to occur in those RR Lyrae stars with greater masses, high luminosities, and longer pulsation periods. However, this conflicts with the proposition of \citet{2020MNRAS.494.1237S}. In their study, the pulsation period is used as the abscissa value. This parameter is not related to the intrinsic parameters of stars, but is related to the evolutionary stage. A long pulsation period does not necessarily mean that the metal abundance of RR Lyrae stars must be low, and a short period does not necessarily mean that the metal abundance is high (see Figure 7 in \citealt{2021ApJ...919..118F}). In addition, in the Bailey diagram, the pulsation amplitude decreases with the pulsation period, which may statistically cause the modulation amplitude to decrease accordingly. The samples studied by \citet{2020MNRAS.494.1237S} are those RRab stars in the Galactic bulge, most of which have pulsation periods of less than 0.7 days. The median photometric metallicity of the whole sample is about -1 dex, which is systematically higher than those Galactic field RR Lyrae stars ([Fe/H] = -1.59, \citealt{2019ApJ...882..169F}). The selection effect cannot therefore be ruled out.

It needs to be acknowledged that the samples in Figure \ref{Fig.12} are still lacking, especially the relatively metal-rich RRab stars ([Fe/H] $>$ -1.0) and stars in globular clusters, as well as the samples of Blazhko RRc and RRd stars. In fact, metal abundance is also the general term for the abundance of elements heavier than hydrogen and helium. It is not expected that this single physical quantity can describe the complex modulation phenomenon, and it cannot be ruled out that other factors might also affect the modulations (e.g. helium and light-element abundances, \citealt{2014MNRAS.443L..15J,2014ApJ...797L...3J}). The RR Lyrae stars in the Galaxy have different formations and evolutions \citep{2022arXiv220804332B}. In the study of their modulation phenomena, stars in different environments and with different properties may need targeted analyses.

Anyway, the potential relationship between metal abundance and Blazhko modulation needs to be further verified from an observational and theoretical perspective. Recently, based on the spectral data of different sky survey projects, the metal abundance of thousands of RR Lyrae stars has been obtained \citep{2019ApJ...882..169F,2020ApJS..247...68L}; The TESS Space Telescope is capable of providing enough photometric data to study the pulsations and modulations that occur in bright RR Lyrae stars. The combination of the two data sources will certainly lead to more comprehensive research results. In terms of theoretical simulations, many models devoted to explaining the Blazhko effect do not consider the influence of different metal abundances, and perhaps new work can be considered from this perspective in the future.

\section{Summary}  \label{Sec:Summary}

Among the 16 Blazhko RRab stars studied by \citet{2013ApJ...773..181N}, V838 Cyg shows the weakest modulation and the highest metal abundance. After noting these characteristics, we made an in-depth study of V838 Cyg using data from several sky surveys, and made a comparative study of the modulations and physical parameters of some Blazhko RRab stars. Finally, we obtained the following results.

1. The $O-C$ diagram shows that the pulsation period of V838 Cyg increases on a long timescale. The specific rate of period change varies depending on whether the early data are used: the rate is 0.050 day Myr$^{-1}$ if used, or 0.211 day Myr$^{-1}$ when they are not considered in the analysis. Further accumulation of observations is needed to obtain a more exact value.

2. Based on the reanalysis of $Kepler$ data, we confirmed the modulation component with a period of 59.45 days discovered in earlier literature, and also found an additional weak modulation with a longer period (about 840 days). The former shows the characteristics of the typical Blazhko effect, while the long-period modulation is simpler, and only shows the changes in pulsation period and amplitude. We suspected the corresponding mechanism is likely to be external (e.g. the LiTE).

3. Based on the modulation and stellar physical parameters provided by \citet{2013ApJ...773..181N}, we found that there is a significant negative correlation between the metal abundance and modulation amplitude of Blazhko RRab stars. We also collected the data from other samples \citep{1994AJ....108.1016L,2012AJ....144...39L}, and found that the metal abundances of the objects with the strongest modulations are poor, and the modulation factor also shows a downward trend with increasing metal abundance. This shows that the metal abundance plays an important role in Blazhko modulation (i.e. the inhibitory effect). This finding may provide some useful support for theoretical models to explain the Blazhko effect.

4. We found that the LC light curves of V838 Cyg show a moir\'{e} pattern that is easy to ignore, and this effect interferes with the relevant research: contrary to the finding of \citet{2014ApJS..213...31B}, V838 Cyg does not show period doubling, and the multiple modulation periods found by \citet{2013ApJ...773..181N} are in fact caused by the moir\'{e} pattern.

When studying the weak modulation of scale from tens of days to hundreds of days, the ultraprecise and uninterrupted photometric data provided by the space telescope are invaluable. Although the accuracy and resolution of the data provided by the TESS space telescope are lower than those of the $Kepler$ space telescope, TESS can still provide a sufficient number of RR Lyrae variable samples for analysis. Combined with the spectroscopic metal abundance data, it is believed that the relationship between Blazhko modulation and metal abundance can be more fully demonstrated.

\acknowledgments

We thank the referee for the valuable comments. This work is supported by the National Natural Science Foundation of China (No.11933008); the Natural Science Foundation of Yunnan Province (No. 202201AT070187, 202301AT070352); the National Natural Science Foundation of China (No. 12103084); and the International Cooperation Projects of the National Key R\&D Program (No.2022YFE0127300). This paper makes use of data from the DR1 of the WASP data (Butters et al. 2010) as provided by the WASP consortium, and computational resources supplied by the project "e-Infrastruktura CZ" (e-INFRA CZ LM2018140) supported by the Ministry of Education, Youth and Sports of the Czech Republic.

Some of the data presented in this paper were obtained from the Mikulski Archive for Space Telescopes (MAST) at the Space Telescope Science Institute. The specific observations analyzed can be accessed via \dataset[https://doi.org/10.17909/T98304]{https://doi.org/10.17909/T98304} and \dataset[https://doi.org/10.17909/3y7c-wa45]{https://doi.org/10.17909/3y7c-wa45}.
STScI is operated by the Association of Universities for Research in Astronomy, Inc., under NASA contract NAS5-26555. Support to MAST for these data is provided by the NASA Office of Space Science via grant NAG5¨C7584 and by other grants and contracts.




%
\clearpage
\begin{table}
\begin{center}
\caption{Times of light maximum and corresponding errors for V838 Cyg provided by present paper. The epoch number and the $O-C$ values are calculated by Equation (\ref{Equ:3}).(The first 10 lines of the whole table.)}\label{Table1}
\begin{tabular}{c c c r c}
\hline\hline
   Max.	       &	Error	  &	  Epoch	    &  $O-C$ & Ref.	\\
HJD-2400000    &	(days)	&		        &	(days)   &      \\
\hline
54231.66773 	&	0.00161 	&	-1526	&	0.00176 	&	SWASP	\\
54232.62931 	&	0.00126 	&	-1524	&	0.00278 	&	SWASP	\\
54256.64342 	&	0.00083 	&	-1474	&	0.00290 	&	SWASP	\\
54257.60145 	&	0.00194 	&	-1472	&	0.00036 	&	SWASP	\\
54280.65827 	&	0.00228 	&	-1424	&	0.00375 	&	SWASP	\\
54282.57607 	&	0.00259 	&	-1420	&	0.00042 	&	SWASP	\\
54283.53631 	&	0.00219 	&	-1418	&	0.00011 	&	SWASP	\\
54284.50162 	&	0.00213 	&	-1416	&	0.00485 	&	SWASP	\\
54609.64631 	&	0.00158 	&	-739	&	0.00005 	&	SWASP	\\
54635.58402 	&	0.00167 	&	-685	&	0.00266 	&	SWASP	\\
... 	&	... 	&	...	&	... 	&	...	\\
\hline\hline
\end{tabular}
\end{center}
\end{table}

\begin{table*}
\centering
\begin{minipage}{175mm}
\caption{The fitted parameters and their estimated errors of the curves of $O-C_{\rm 1, max}$, $O-C_{\rm 1, min}$, pulsation amplitudes (Full Amp., $A_{1}$, $Mag_{\rm max}$) and Fourier coefficients ($R_{21}$, $R_{31}$, $\phi_{21}$, and $\phi_{31}$) curves in Figure~\ref{Fig.2}. The software was used for analysis is Period04 \citep{2005CoAst.146...53L}.}\label{Table2}
{\scriptsize
\def\arraystretch{1.5}
\tabcolsep=1.3pt
\begin{tabular}{l c c c c c c}
\hline\hline
	            &	$f_{\rm S}$	&	$A_{\rm S}$	&$\phi_{\rm S}$	&	$f_{\rm L}$	&	$A_{\rm L}$	&	$\phi_{\rm L}$	\\
	            &	(day$^{-1}$)	&		        &	(deg)	&	(day$^{-1}$)	&		        &	(deg)	\\
\hline
$O-C_{\rm 1, max}$ (days)	&	0.01685(3)	&	0.00009(1)	&	2(5)	&	0.00122(2)	&	0.00013(1)	&	213(3)	\\
$O-C_{\rm 1, min}$ (days)	&	0.01685(4)	&	0.00014(1)	&	272(5)	&	0.00121(4)	&	0.00014(1)	&	217(5)	\\
Full amp. (mag)	&	0.01683(4)	&	0.00178(16)	&	46(5)	&	0.00124(2)	&	0.00308(16)	&	151(3)	\\
$A_{1}$ (mag)	    &	0.01681(1)	&	0.00202(5)	&	347(1)	&	0.00122(2)	&	0.00100(5)	&	155(3)	\\
$Mag_{\rm max}$ (mag)	&	0.01679(3)	&	0.00150(11)	&	15(4)	&	0.00105(2)	&	0.00165(11)	&	156(4)	\\
$R_{21}$	    &	0.01682(1)	&	0.00217(5)	&	121(1)	&	\-	&	\-	&	\-	\\
$R_{31}$ 	    &	0.01682(1)	&	0.00307(4)	&	140(1)	&	\-	&	\-	&	\-	\\
$\phi_{21}$	(rad) &	0.01677(2)	&	0.00181(9)	&	38(3)	&	\-	&	\-	&	\-	\\
$\phi_{31}$ (rad) &	0.01683(1)	&	0.00532(12)	&	193(1)	&	\-	&	\-	&	\-	\\
\hline
Mean frequency    & 0.01682(2)  &    \-         &     \-    &   0.00119(3)    &   \-   &   \-  \\
Mean period       & 59.45(7)    &    \-         &     \-    &   840(21)       &   \-   &   \-  \\
\hline\hline
\end{tabular}
}
\end{minipage}
{\scriptsize {\bf Notes.} The letters S and L represent the short-period modulation and the long-period modulation, respectively. The digits in parentheses are the errors in the final digits of the corresponding quantities. For example, 0.00122(2) means 0.00122$\pm$0.00002, and 213(3) means 213$\pm$3.}
\end{table*}

\begin{table*}
\begin{center}
\caption{The pulsation and orbital elements of V838 Cyg. The descriptions of the orbital elements and corresponding fitting equations can be seen in \citet{2014MNRAS.444..600L}.}\label{Table3}
\begin{tabular}{c c c}
\hline\hline
Parameter                  &  V838 Cyg (MAX)               & V838 Cyg (MIN)\\
\hline
$T_{0}[\rm cor]$           & $2454964.57632(2)$            & $2454964.51459(3)$\\
$P_{0}[\rm cor]$           & $0.48027930(3)$               & $0.48027928(5)$\\
$\beta$ (d Myr$^{-1}$)     & $0.34\pm0.01$                 & $0.36\pm0.02$\\
$A (\rm s)$                & $15.1\pm1.9$                  & $23.7\pm9.3$\\
$a_{1}\sin i (\rm AU)$     & $0.030\pm0.004$               & $0.047\pm0.019$\\
$e$                        & $0.65\pm0.10$                 & $0.82\pm0.15$\\
$\omega$ (deg)             & $4.7\pm7.7$                  & $357.4\pm7.1$\\
$P_{\rm orb}$ (day)          & $836.3\pm14.2$                & $866.8\pm18.1$\\
$T$                        & $2455183.9\pm36.2$            & $2455149.1\pm45.1$\\
\hline\hline
\end{tabular}
\end{center}
\end{table*}

\begin{table*}
\centering
\begin{minipage}{175mm}
\caption{Information on RR Lyrae stars that show moir\'{e} patterns in their $Kepler$ light curves.}\label{Table1App}
\begin{tabular}{c c c c c c}
\hline\hline
Star Name  & $T_{0}$     & $P_{0}$      & $P_{\rm pul}/\Delta t_{\rm LC}$ &  $f_{\rm beat}$ & $f_{\rm b1,1}$,$f_{\rm b1,2}$ \\
           &(BJD-2400000)&   (days)     &                        &    (day$^{-1}$)   &              (day$^{-1}$)          \\
\hline
Type I     &             &              &          &                       &                          \\
V715 Cyg   &54964.60622  &0.47070588    & 23.03587 & 0.07629(2)           &0.0762*,2.0483            \\
V784 Cyg   &54964.80988  &0.53409458    & 26.13805 & 0.25855(1)           &0.2585*,1.6138            \\
KIC 6100702 &54953.84603  &0.48814517    & 23.88934 & 0.226655(3)          &1.8219,0.2267*            \\
V350 Lyr   &54964.78382  &0.59423786    & 29.08141 & 0.137023(1)          &0.1370*,1.5458            \\
V894 Cyg   &54953.56580  &0.57138644    & 27.96309 & 0.06453(1)           &1.6855,0.0646*            \\
KIC 9658012 &55779.94646  &0.53319459    & 26.09401 & 0.17648(3)           &0.1763*,1.6992            \\
\hline
Type II    &             &              &          &                       &                          \\
V838 Cyg   &54964.57594  &0.48028001    & 23.50442 & 0.01684(3)           &1.0503,1.0319             \\
           &             &              &          &                &0.0184(=$f_{\rm b1,1}-f_{\rm b1,2}$)  \\
\hline
Type III   &             &              &          &                       &                          \\
V368 Lyr   &54964.78830  &0.45648601    & 22.33997 & -                     &0.7448,1.4459             \\
           &             &              &          &                &0.0436(=$2f_{\rm b1,1}-f_{\rm b1,2}$)   \\
\hline\hline
\end{tabular}
\end{minipage}
{\scriptsize {\bf Notes.} The columns contain: (1) star name; (2) time of light maximum (BJD-2400000); (3) pulsation period in days; (4) the ratio of the pulsation period and the LC sampling period; (5) beat frequencies obtained from the Fourier spectrum of $O-C$ diagrams; (6) beat frequencies calculated from Equations. (\ref{Equ:4}, \ref{Equ:5}, \ref{Equ:6}, \ref{Equ:7}). In type I, the beat frequencies from the spectrum of $O-C$ diagrams are approximately equal to the smaller of the two values in column 6 (labeled with asterisks).}
\end{table*}


\clearpage



\end{document}